\documentclass[sigconf]{acmart}

\usepackage{soul}
\usepackage{parskip}
\usepackage{graphicx}
\usepackage{listings}
\usepackage{xcolor}

\definecolor{codegreen}{rgb}{0,0.6,0}
\definecolor{codegray}{rgb}{0.5,0.5,0.5}
\definecolor{codepurple}{rgb}{0.58,0,0.82}
\definecolor{backcolour}{rgb}{0.95,0.95,0.92}
\lstdefinestyle{mystyle}{
    backgroundcolor=\color{backcolour},   
    commentstyle=\color{codegreen},
    keywordstyle=\color{magenta},
    numberstyle=\tiny\color{codegray},
    stringstyle=\color{codepurple},
    basicstyle=\ttfamily\footnotesize,
    breakatwhitespace=false,         
    breaklines=true,                 
    captionpos=b,                    
    keepspaces=true,                 
    numbers=left,                    
    numbersep=5pt,                  
    showspaces=false,                
    showstringspaces=false,
    showtabs=false,                  
    tabsize=2,
}

\AtBeginDocument{%
  }

\setcopyright{acmlicensed}
\copyrightyear{2018}
\acmYear{2018}
\acmDOI{XXXXXXX.XXXXXXX}

\acmConference[Conference acronym 'XX]{Make sure to enter the correct
  conference title from your rights confirmation emai}{June 03--05,
  2018}{Woodstock, NY}
\acmISBN{978-1-4503-XXXX-X/18/06}

\begin{document}

\title{Using a Feedback Loop for LLM-based Infrastructure as Code Generation}

\author{Mayur Amarnath Palavalli}
\email{mayur.palavalli@gmail.com}
\affiliation{%
  \institution{Irvington High School}
  \city{Fremont}
  \state{California}
  \country{USA} 
}

\author{Mark Santolucito}
\email{msantolu@barnard.edu}
\affiliation{%
 \institution{Barnard College, Columbia University}
  \city{New York}
  \state{New York}
  \country{USA} 
 }

\renewcommand{\shortauthors}{Trovato et al.}

\begin{abstract}
Code generation with Large Language Models (LLMs) has helped to increase software developer productivity in coding tasks, but has yet to have significant impact on the tasks of software developers that surround this code.
In particular, the challenge of infrastructure management remains an open question.
We investigate the ability of an LLM agent to construct infrastructure using the Infrastructure as Code (IaC) paradigm.
We particularly investigate the use of a feedback loop that returns errors and warnings on the generated IaC to allow the LLM agent to improve the code.
We find that, for each iteration of the loop, its effectiveness decreases exponentially until it plateaus at a certain point and becomes ineffective.  
\end{abstract}



\received{20 February 2007}
\received[revised]{12 March 2009}
\received[accepted]{5 June 2009}

\maketitle

\section{Introduction}
Infrastructure as Code (IaC) has fundamentally transformed the way cloud infrastructure is managed. 
With IaC, developers can provision and maintain their infrastructure through code. 
This ensures automation and consistency in deploying infrastructure, allowing for more effective scaling of operations. 
IaC also allows teams to use version control on their infrastructure, making it easier to collaborate and track changes~\cite{8919181}. 
However, a major challenge with IaC is in the difficulty of writing correct code~\cite{10.1007/978-3-030-72013-1_6,10174011,9779848,cauli2022pre,qiao2024statically}.

At the same time, Large Language Models (LLMs), as applied to code generation, are enabling developers to be more effective.
Code generation benchmarks, such as HumanEval~\cite{githubGitHubOpenaihumaneval} and SWEBench~\cite{githubGitHubPrincetonnlpSWEbench}, have shown that LLMs are capable of assisting developers with challenging programming tasks.
It is then natural to seek to extend the application of LLMs for code generation to IaC.

The combination of IaC and LLMs could allow for wider adoption of IaC and more effective infrastructure management. 
However, many questions remain open about the ability of LLMs to reason about the complexities of IaC.
In particular, there are many implicit rules about the semantics of cloud resources that are difficult to reconcile when creating infrastructure through IaC.

In our work, we provide an analysis of the ability of LLMs to generate AWS CloudFormation code.
In particular, we investigate the ability of an LLM to respond to errors and warnings generated by \texttt{cfn-lint}~\cite{githubGitHubAwscloudformationcfnlint}, a tool for analyzing CloudFormation code.
In summary, the main contributions of this work are:
\begin{enumerate}
    \item The design of a feedback loop system for the generation of CloudFormation code.
    \item A set of CloudFormation code generation benchmarks built from industry-standard IaC problems.
    \item An evaluation of our feedback loop system on this benchmark, showing that LLMs struggle to fully reconcile all errors in IaC generation.
\end{enumerate}

\section{Related Work}

The application of LLMs in generating IaC has amassed significant attention in recent years. Using LLMs for IaC comes with certain challenges that make it inefficient in large systems. Srivatsa et al. evaluate the LLM performance on functional correctness by comparing with human-written code and deciding whether or not it is an exact match. They found that the GPT-3.5-turbo model had a success rate of between 50 and 60 percent, while the Codeparrot model never exceeded 10 percent accuracy~\cite{srivatsa2024survey}. They also discuss ethical and safety concerns of developing LLMs for more accurate IaC generation.  

A natural extension is the work of Ugare et al. who introduce SynCode~\cite{ugare2024syncodellmgenerationgrammar}, a framework that uses grammar rules to enhance LLM generation in formal coding languages. SynCode is able to reduce 96.07 percent of syntax errors in Python and Go. It particularly shines in generating JSON, where it is able to eliminate all syntax errors. This is achieved by utilizing context-free grammar rules based on discrete finite automation. 

An important area of investigation related to LLM code generation is constrained decoding~\cite{park2024grammaraligneddecoding, arxivGrammarConstrainedDecoding}. 
At a high level, grammar constrained decoding (GCD) helps language models (LLMs) in producing structured results without requiring additional fine tuning. GCD utilizes grammars to guarantee that the generated sequences follow a predefined structure. 
The technique greatly improves the performance of LLMs in such settings without the need for expensive task-specific training efforts.
Since we are generating highly structured JSON documents (CloudFormation files), GCD could be a way to further optimize the results. 
However, the main goal of our work is to provide an initial baseline accounting of the viability of LLM code generation for IaC.
We leave such optimizations to future work.

Another critical challenge in LLM code generation is fixing syntax errors. Tsai et al. address this with RTLFixer~\cite{tsai2023rtlfixer}, a framework designed to fix syntax errors in Verilog code.
After finding that 55 percent of errors in LLM-generated Verilog code were syntax errors, RTLFixer was designed to utilize Retrieval-Augmented Generation and ReAct (Reasoning and Action framework) techniques to improve error correction. The framework achieves a 98.5 percent success rate in fixing syntax errors after testing on 212 syntactically invalid Verilog implementations. 

These studies demonstrate the evolving role of LLMs in code generation and highlight the importance of addressing syntactic correctness and complexity.


\section{Background}
\begin{figure}[h!]
    \centering
\begin{lstlisting}
{"AWSTemplateFormatVersion": "2010-09-09",
"Resources": {
  "MyEC2Instance": {
    "Type": "AWS::EC2::Instance",
    "Properties": {
      "InstanceType": "t2.micro",
      ...
\end{lstlisting}
    \caption{An example AWS CloudFormation JSON template.}
    \label{fig:iac}
\end{figure}

IaC is a diverse space with many existing languages and tools.
The most widely adopted tools are Terraform~\cite{githubGitHubHashicorpterraform} and Amazon Web Service's CloudFormation~\cite{amazonWhatCloudFormation}.
Both Terraform and CloudFormation files are specified as JSON (or YAML) documents, where the various fields define properties of the cloud resource to be deployed.
These files are generally declarative - giving a specification of the desired cloud infrastructure state.
This is in contrast to some other IaC languages, such as Pulumi, which allow users to write imperative programs that construct operations that should take place upon the user's infrastructure~\cite{pulumiDocumentation}.
An example of a CloudFormation JSON is given in Fig.~\ref{fig:iac}.

\begin{figure}[h!]
    \centering
\begin{lstlisting}
E1015 {'Fn::GetAZs': ''} is not of type 'string'
Error location - path/to/my_iac.json:1:4575
\end{lstlisting}
    \caption{An example error message from \texttt{cfn-lint}.}
    \label{fig:error}
\end{figure}

In this work, we focus solely on the LLM generation of AWS CloudFormation~\cite{amazonWhatCloudFormation} in JSON.
We additionally use the AWS CloudFormation linter \texttt{cfn-lint}~\cite{githubGitHubGofireflyioaiac}.
\texttt{cfn-lint} allows us to validate CloudFormation JSON templates against the resource provider schemas provided by AWS as well as other best-practice IaC rules. 
A schematically valid JSON template returns nothing when run through \texttt{cfn-lint}. 
In any other case, \texttt{cfn-lint} returns errors and/or warnings. 
The example error in \textbf{Fig.~\ref{fig:error}} outlines the basic structure of an error produced by \texttt{cfn-lint}.
The error message code starts with  a letter followed by a string of numbers which forms a error code. 
This is followed by a brief description of the error and the line and character number at which it occurs.


\section{Methodology}
To describe our system, we first describe the methodology we used for collecting a benchmark set, then we describe the feedback loop that we constructed with \texttt{cfn-lint}.
We make all of our code and evaluation results available open-source: \url{https://github.com/Mayur-Palavalli/LLM-IaC-generation}.

\begin{figure}[h!]
    \centering
\begin{lstlisting}[breakindent=0pt]
Create a AWS CloudFormation template that deploys a VPC with a pair of private subnets spread across two Availabilty Zones. It deploys a VPC Endpoint for CloudFormation so an instance in the private subnet can use cfn-signal for its CreationPolicy.
\end{lstlisting}
    \caption{An example prompt from the official AWS CloudFormation Template Schema repository.}
    \label{fig:prompt}
\end{figure}

\subsection{Benchmark Set}
To create a dataset of prompts, we took 33 descriptions of AWS CloudFormation problems from the official AWS CloudFormation Template Schema repository~\cite{githubGitHubAwscloudformationawscloudformationtemplates}.
This repository converts existing Resource Specifications files into a formatted JSON (or YAML) schema document which can be integrated in an IDE. An example prompt from this repository is displayed in Fig.~\ref{fig:prompt}

\subsection{Feeback Loop}
We queried an LLM for a solution to this problem five times for each description, yielding a dataset of 165 implemented CloudFormation templates.
We generated these templates in JSON format.

\begin{figure}
    \centering
    \includegraphics[width=8cm]{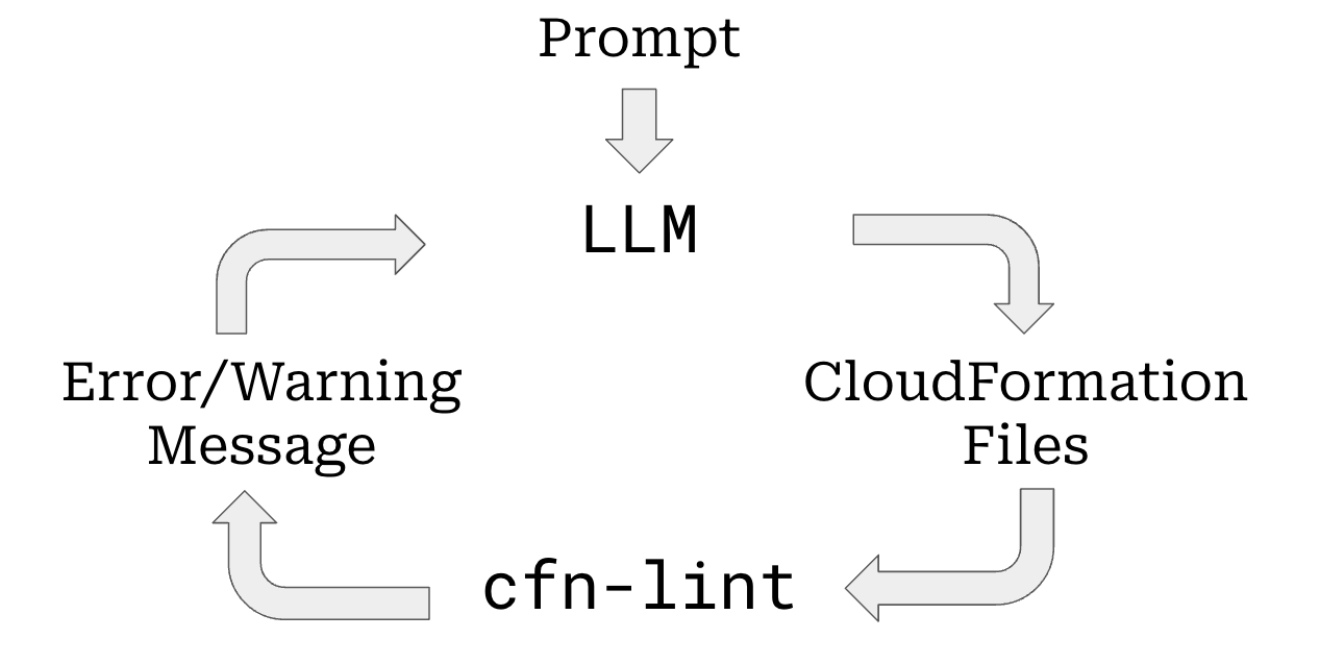}
    \caption{A diagram of the feedback loop: we provide the LLM with a prompt for an AWS CloudFormation file, which is run through \texttt{cfn-lint} to produce error/warning message(s) which are given back to the LLM.}
    \label{fig:feedbackLoop}
\end{figure}

The tool \texttt{cfn-lint} is able to process a JSON CloudFormation template, and returns errors describing schema violations, invalid resource properties, and best practices.
After each generation, we run \texttt{cfn-lint}, and send the error message(s) and JSON template back to the LLM, instructing it to modify the template to fix the error generated by \texttt{cfn-lint}.
In this manner, we create the feedback loop outlined in \textbf{Fig.~\ref{fig:feedbackLoop}} that sends each JSON CloudFormation file in our dataset to the LLM, each time providing the new file and corresponding error message(s). 
This process is repeated ten times.
We hypothesized that the LLM would be able to fix all errors produced by \texttt{cfn-lint} after a certain number of iterations.

We keep track of the total number of errors and warnings across all 165 files after each iteration of the feedback loop (for all 10 iterations) in order to identify the point at which it becomes ineffective. 
This process of running each file ten times through the feedback loop is repeated six times. 
We choose to repeat it six times because most of the error bars are narrow, indicating a small standard deviation in the number of errors after each of the ten iterations.
A small standard deviation implies low variance, which demonstrates that additional repetitions are unlikely to change the overall pattern. 

\section{Results}

\begin{figure}[htp]
    \centering
    \includegraphics[width=8cm]{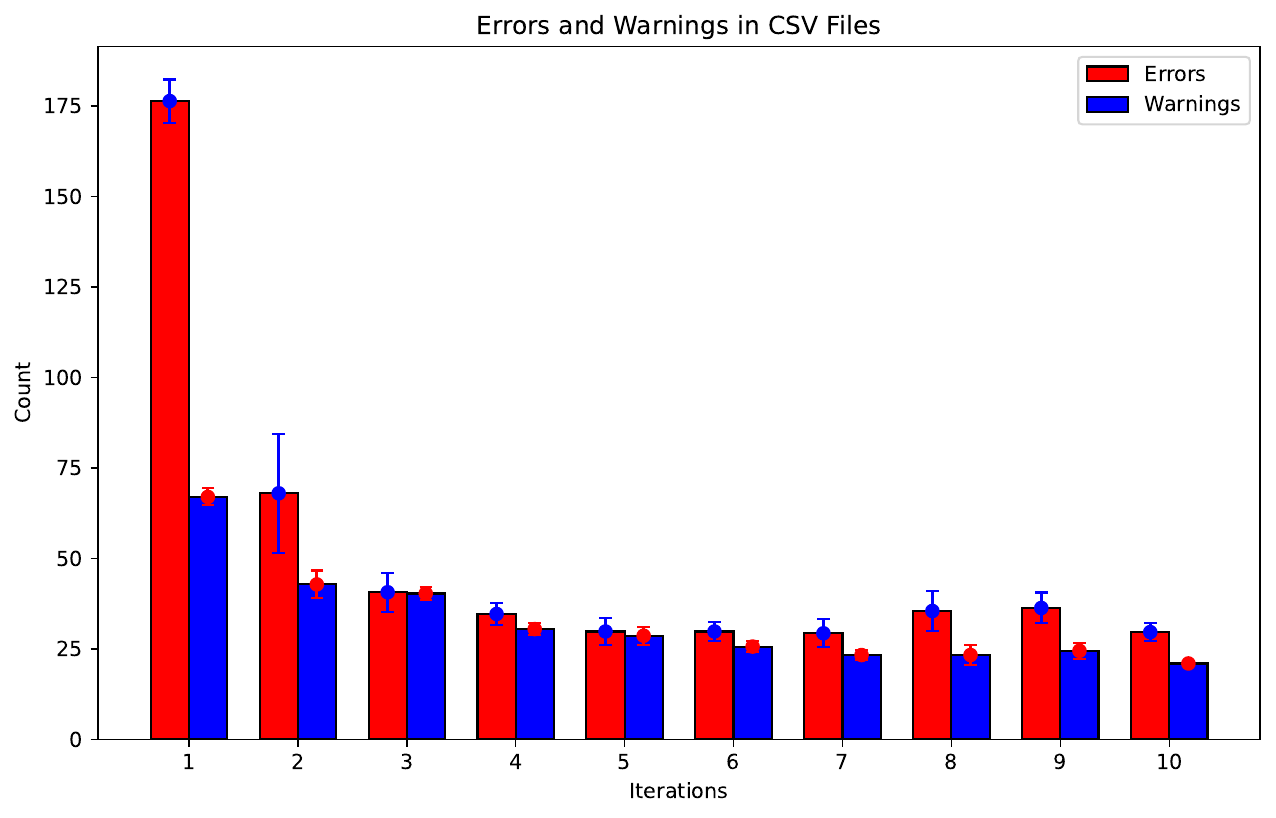}
    \caption{A histogram of errors over multiple \texttt{cfn-lint} feedback iterations showing error bars representing the distribution over six trials.}
    \label{fig:errorChart}
\end{figure}

The entire process of iterating each file ten times through the feedback loop is repeated six times. \textbf{Fig.~\ref{fig:errorChart}} summarizes the results of all six sets of iterations in a bar graph by displaying the total number of errors and warnings in all 165 JSON files after each iteration. The graph's plateau beginning at approximately the fifth iteration indicates the point at which the feedback loop is no longer effective. 

The plateau is caused by the feedback loop's inconsistency in fixing errors. 
The LLM is unable to correctly fix certain individual errors, which occasionally result in several new errors being produced from each old error between iterations. 
After approximately five iterations, this anomaly is enough to yield no significant net change in the number of errors across all 165 files. 
The slight peak in the number of errors at the eighth and ninth iterations indicates at least one of two things: 
\begin{enumerate}
    \item An unusually high number of files generated more errors than it had after the seventh iteration. 
    \item Certain files had an unusually high increase in the number of errors generated after the seventh iteration.
\end{enumerate}
We predict the cause of this to be the LLMs incapability to properly understand all the error messages from \texttt{cfn-lint}.

\section{Discussion}

Perfecting the use of LLMs in generating valid AWS CloudFormation could enable developers to more quickly build the infrastructure they need for their systems.
Automating the generation of CloudFormation templates could drastically increases the speed and efficiency of setting up complex cloud environments.  

We do not use ChatGPT's structured output mode~\cite{OpenAIstructuredoutputs}. 
Structured output mode allows users to provide the desired schema of JSON output from ChatGPT in addition to the prompt.
This mode guarantees that the generated response matches the schema.
However, for the purposes of CloudFormation generation this is not a viable option.
Not only is the complete CloudFormation schema is over 200,000 lines long, it uses features of JSON schemas that are outside the scope of ChatGPT's structured output mode~\cite{OpenAIstructuredoutputs}.

While our feedback loop dramatically increases the chance of generating valid IaC, we still do not have a formal guarantee that the generated code will be schematically valid. 
The remaining uncertainty is enough for LLM code generation of IaC to remain hard to use in large-scale development. 
There is a further question of semantic validity. 
Semantic validity captures the idea that not only is the CloudFormation deployable, but is also what the user wants (e.g. an empty file is always schematically valid, but not semantically valid). 
A schematically valid CloudFormation is not guaranteed to be semantically valid.
We are left with the task of ensuring schematically valid CloudFormation are also semantically valid before LLMs are safe to use at large scale. 
This, however, is a challenging problem.
Measuring semantic validity requires a way to determine how well the generated infrastructure adheres to the prompt.
There is currently no simple tool like \texttt{cfn-lint} that does this.

\subsection{Threats to Validity}
One threat to the generalizability of our work is the question of the extent to which we would see the same pattern if using a different LLM.
At the time of writing, OpenAI's gpt-4o is among the best LLMs capable of working with JSON.
We believe that the use of a different LLM for our work would the rate of decrease in the number of errors, but not change the overall pattern. 
We hypothesize that the plateau on the graph in Fig.~\ref{fig:errorChart} is due to the LLM's inability to reconcile error messages with the high-level intent of the prompt.
If this is true, we would need a structurally different LLM than OpenAI's gpt-4o model.

Another threat to generalizability is the extent to which the IaC provider impacts the pattern of our results. 
AWS CloudFormation is one of the most well-documented IaC services, and thus has large corpus of training data related to it, making it well-suited to generation with LLMs.
Other well-documented IaC services, such as Terraform, would likely yield similar results due to similar compatibility with LLMs. 
We hypothesize that using a less documented services would likely still produce a plateau, but would yield more slowly declining error bars due to a smaller training set of relevant data. 

Furthermore, our work utilizes only one method of receiving feedback. 
We use error messages from \texttt{cfn-lint} as feedback to demonstrate a proof of concept, but other means of testing and receiving feedback could yield different results. 
The plateau, for instance, might not exist with a more structured feedback strategy. 

\subsection{Future Work}
We find that an LLM has a limited ability to respond to error messages and correct code in the context of IaC configurations.
A future task would be to customize error messages to a format that an LLM like ChatGPT more clearly understands so errors can be be fixed without creating new ones. 
The creation of a tool like cfn-lint to check for semantic validity would take IaC generation via LLMs to the next level, allowing for widespread commercial use. 


In the context of IaC, Pulumi~\cite{pulumiDocumentation} is a an open-source IaC tool that allows developers to define, deploy, and manage cloud infrastructure in familiar programming languages. 
All AWS services, including CloudFormation, are fully supported by Pulumi. 
An especially useful tool is Pulumi AI~\cite{githubGitHubPulumipulumiai}, an experimental feature that allows developers to generate IaC in familiar programming languages via natural-language prompts. 
As far as we know, the Pulumi AI tool does not incorporate any feedback during its generation from static analysis tools such as \texttt{cfn-lint}.
A future task might be to use Pulumi AI to generate Pulumi code which maps to AWS CloudFormation, which is run through  \texttt{cfn-lint}, and send the generated errors back to correct the Pulumi code.

\section{Conclusion}
Our investigation of the use of feedback loop for Infrastructure as Code generation demonstrates the potential of agentic LLM systems for real-world software development.
Our results indicate that using LLMs to generate IaC provides a significant advantage, but not yet enough to be make it scale-able. 
The feedback loop offers a method to increase the rate of successful IaC generation from LLMs, as well as an opportunity for further research to fix the anomalies that cause the loop to become eventually ineffective. 
Even if OpenAI's structured output mode were to support large complex schema like that of AWS CloudFormation, there are still issues of semantic correctness that are not captured in the schema and only appear through checks with tools like \texttt{cfn-lint}. 

\bibliographystyle{splncs04}
\bibliography{references}

\end{document}